\documentclass[12pt]{article}
\usepackage{amsmath,amssymb,amscd,cite}
\newtheorem{thm}{Theorem}[section]
\newtheorem{prop}[thm]{Proposition}
\newtheorem{lem}[thm]{Lemma}
\newtheorem{cor}[thm]{Corollary}

\newtheorem{defi}[thm]{Definition}
\newcommand{\pf}{{\bf Proof. \ }}
\newcommand{\qed}{\hfill $\Box$ \\}
\font\msbm=msbm10 at 12pt
\newcommand{\Z}{\mbox{\msbm Z}}

\newcommand{\F}{\mbox{\msbm F}}
\newtheorem{rem}[thm]{Remark}
\newtheorem{ex}[thm]{Example}

\begin{document}
\title{Repeated Root Constacyclic Codes of Length $mp^s$ over $\mathbb{F}_{p^r}+u \mathbb{F}_{p^r}+\ldots
+ u^{e-1}\mathbb{F}_{p^r}$}
\author{Kenza Guenda and T. Aaron Gulliver
\thanks{ T. Aaron Gulliver is
with the Department of Electrical and Computer Engineering,
University of Victoria, PO Box 3055, STN CSC, Victoria, BC, Canada
V8W 3P6. email: agullive@ece.uvic.ca.}}
 \maketitle
\begin{abstract}
We give the structure of $\lambda$-constacyclic codes of length
$p^sm$ over $R=\mathbb{F}_{p^r}+u \mathbb{F}_{p^r}+\ldots +
u^{e-1}\mathbb{F}_{p^r}$ with $\lambda \in \F_{p^r}^*$. We also give
 the structure of $\lambda$-constacyclic codes of length $p^sm$ with
$\lambda=\alpha_1+u\alpha_2+\ldots +u^{e-1} \alpha_{e-1}$, where
$\alpha_1,\alpha_2 \neq 0$ and study the self-duality of these codes.
\end{abstract}

\section{Introduction}
Codes over rings were introduced by Blake~\cite{blake} as a generalization of codes over fields.
It has been shown~\cite{sole1} that these codes are related to some of the best known non-linear codes,
such as the Kerdock, Preparata and Goethals codes.
Subsequently, finite rings such as finite chain rings~\cite{CS,Ana,permounth} and finite principal
rings~\cite{CRT,G-G} have been used to construct codes.
Codes over the polynomial residue rings $\frac{\F_{p^r}[u]}{u^{e}}=\F_{p^r}+u\F_{p^r}+\ldots+u^{e}\F_{p^r}$
were introduced by Bachoc~\cite{bachoc}.
Many properties such as the code structure, minimum distance, self-duality and decoding have been studied by
Bonnecaze and Udaya~\cite{udaya1}, Dougherty et al.~\cite{type2}, and Gulliver and Harada~\cite{gulliver1,gulliver2}.
Codes over these rings have been used in the construction of codes for DNA
computing by Siap et al.~\cite{siap}, and more recently by Abualrub and
Siap~\cite{abualrub} and the authors~\cite{G-G12}.
These rings are a special case of finite chain rings.

Simple root constacyclic codes over finite chain rings are well known~\cite{permounth,G-G}.
It was proven by S\u al\u agean~\cite{salagean} that if $(n,p)\neq 1$, cyclic repeated root
codes of length $n$ over finite chain rings are not principal ideals. This is also
true for negacyclic codes if $p$ is odd. However, little is known
about repeated root constacyclic codes over finite chain rings, or
even over polynomial residue rings. Recently, the structure of
some constacyclic codes over polynomial residue rings of length
$p^s$ was given by Jitman and Udomkanavich~\cite{jitman}. Dinh and
Nguyen also studied some classes of repeated root constacyclic codes
over polynomial rings of even characteristic and constacyclic codes
of length $p^s$ over $\F_{p^r}+u\F_{p^r}$~\cite{dinh}.
In the first part of this paper, we give the structure of $\lambda$-constacyclic
codes of length $p^sm$ over $\mathbb{F}_{p^r}+u
\mathbb{F}_{p^r}+\ldots + u^{e-1}\mathbb{F}_{p^r}$, where $\lambda
\in \F_{p^r}^*$. Furthermore, we prove that some of these
$\lambda$-constacyclic codes are equivalent to cyclic codes. In the
second part of this paper we deal with $\lambda$-constacyclic
codes of length $p^sm$ over $\mathbb{F}_{p^r}+u
\mathbb{F}_{p^r}+\ldots + u^{e-1}\mathbb{F}_{p^r}$ with
$\lambda=\alpha_1+u\alpha_2+\ldots +u^{e-1} \alpha_{e-1}$, where
$\alpha_1,\alpha_2 \neq 0$.
In addition, the self-duality of these codes is considered.
This extends some results in~\cite{dinh}.

The remainder of this paper is organized as follows. Section 2 gives
some preliminary results concerning finite chain rings. In Section
3, we study $\lambda$-constacyclic codes of length $p^sm$ over
$R=\mathbb{F}_{p^r}+u \mathbb{F}_{p^r}+\ldots +
u^{e-1}\mathbb{F}_{p^r}$ when $\lambda$ is a $p^s$ power in
$\F_{p^r}$. The equivalence of these codes to cyclic codes over $R$
is established. In Section 4, we study $\lambda$-constacyclic codes
of length $p^sm$ over $R=\mathbb{F}_{p^r}+u \mathbb{F}_{p^r}+\ldots
+ u^{e-1}\mathbb{F}_{p^r}$, where $\lambda=\alpha_1+u\alpha_2+\ldots
+u^{e-1}\alpha_{e-1}$ such that $\alpha_1,\alpha_2 \neq 0$. It is
proven that these codes are principally generated, and the
self-duality of these codes is considered.

\section{Preliminaries}

Recall that a finite chain ring is a finite local, principal
commutative ring $R$ with $1\neq 0$ such that its ideals are ordered
by inclusion
\[
<0>= \langle u^e\rangle \subsetneq \langle u^{e-1}\rangle \subsetneq \ldots \subsetneq \langle u\rangle \subsetneq R.
\]
The residue field of the finite chain ring $R$ is
$\F_{p^r}= R/\langle u \rangle .$
The polynomial residue ring over $\F_{p^r}$ with nilpotency index $e$ is the finite chain ring
$R=\frac{\F_{p^r}[u]}{u^{e}}=\mathbb{F}_{p^r}+u\mathbb{F}_{p^r}+\ldots + u^{e-1}\mathbb{F}_{p^r}$, where $u^{e}=0$.
This ring has nilpotency index $e$ and a unique maximal ideal $\langle u \rangle $.
An element $v\in R$ is uniquely expressed as
\[
r=v_0+uv_1+\ldots u^{e-1}v_{e-1},
\]
with $r_i \in F_{p^r}$.
The units of $R$ are of the form
$v=v_0+uv_1+\ldots +u^{e-1}v_{e-1}$ with $v_0 \neq 0$.
Hence if the group of units is denoted as $R^*$ we have $R^*=R\setminus \langle u \rangle$.

A code $\mathcal{C}$ of length $n$ over $R$ is a subset of $R$.
If the code is a submodule we say that the code is linear.
For a given unit $\lambda \in R$, a code $\mathcal{C}$ is said to be constacyclic, or more generally,
$\lambda$-constacyclic, if $(\lambda c_{n-1}, c_0, c_1, \ldots, c_{n-2})\in \mathcal{C}$ whenever
$(c_0, c_1,\ldots, c_{n-1}) \in \mathcal{C}$.
For example, cyclic and negacyclic codes correspond to $\lambda=1$ and $-1$, respectively.
It is well known that the $\lambda$-constacyclic codes over a finite chain ring $R$ correspond
to ideals in $R[x]/\langle x^n-\lambda \rangle$, which in this paper is denoted as $\mathcal{R}$.
The dual code $\mathcal{C}^\perp$ of $\mathcal{C}$ is defined as
\begin{equation}
\mathcal{C}^\perp=\{ {v} \in R^n \  | \ [{v},{w}]= 0 \mbox{ for all } w \in \mathcal{C}\}.
\end{equation}
The dual of a $\lambda$-constacyclic code is a $\lambda^{-1}$-constacyclic code.

For a polynomial $f(x)=a_0+a_1 x + \ldots+ a_rx^r$ with $a_0 \neq 0$
and degree $r$ (hence $a_r\neq 0$), the reciprocal of $f$ is the polynomial
\begin{equation}
\label{eq:5} f^*(x)=x^rf(x^{-1})=a_r+a_{r-1}x+\ldots + a_0x^r.
\end{equation}
If a polynomial $f$ is equal to its reciprocal, then $f$ is called
self-reciprocal. We can easily verify the following equalities
\begin{equation}
\label{eq:prop}
 (f^*)^*=f \mbox{ and } (fg)^*=f^*g^*.
\end{equation}
Let $S$ be a nonempty set of $R$, then the annihilator of $S$ denoted by $ann(S)$ is the set
\[
ann(S)=\{f|fg=0 \text{ for all } g \in S\}.
\]

\begin{lem}(\cite[Proposition 3.4]{dinh})
\label{sel:dinh} Let $R$ be a commutative ring and $\lambda$ a unit in
$R$ such that $\lambda^2=1$. If $\mathcal{C}$ is a
$\lambda$-constacyclic code over $R$, then the dual code
$\mathcal{C}^{\bot}$ of $\mathcal{C}$ is the ideal
$ann^*(\mathcal{C})$ where $ann^*(\mathcal{C})=\{f^*(x) |
f(x)^*g(x)= 0 \text{ for all } g(x) \in \mathcal{C}\}$.
\end{lem}

If $\mathcal{C} = \mathcal{C}^\perp$, we say that the code is self-dual.
A code $\mathcal{C}$ and its dual satisfy the following
\begin{equation}
\label{wood}
|\mathcal{C}||\mathcal{C}^{\bot}|=p^{ren}=|R|^n \mbox{ and }(\mathcal{C}^{\bot})^{\bot}=\mathcal{C}.
\end{equation}
\begin{rem}
\label{rem:wood}
From (\ref{wood}), there exists a self-dual code of length $n$ over $R$ if and only if $en$ is even.
\end{rem}
\begin{defi}
A polynomial $f$ in $R[x]$ is called regular if $\overline{f} \neq 0$.
$f$ is called primary if the ideal $\langle f \rangle$ is
primary, and $f$ is called basic irreducible if $\overline{f}$ is irreducible in $F_{p^r}[x]$.
Two polynomials $f$ and $g$ in $R[x]$ are called coprime if
\[
R[x]=\langle f \rangle +\langle g \rangle.
\]
\end{defi}

\begin{lem}(\cite[Theorem XIII. 11]{Mac})
\label{XIIIMac} Let $f$ be a regular polynomial in $R[x]$.
Then $f=\alpha g_1 \ldots g_r$, where $\alpha$ is a unit and $g_1, \ldots, g_r$ are regular primary coprime polynomials.
Moreover, $g_1, \ldots , g_r$ are unique in the sense that if $f=\alpha g_1
\ldots g_r=\beta h_1 \ldots h_s$, where $\alpha, \beta$ are units
and $g_i$ and $h_i$ are regular primary coprime polynomials, then
$r=s$, and after renumbering $\langle g_i \rangle= \langle h_i \rangle$, $1\le i \le n$.
\end{lem}

Several weights over rings can be defined.
The homogenous weight is defined~\cite{greferath2} as the following generalization of the Lee weight
\begin{itemize}
\item[(i)] $\forall x\in R \setminus \langle u^{e-1} \rangle$, then $w(x)=p^{r(e-2)}(p^r-1)$;
\item[(ii)]  $\forall x\in  \langle u^{e-1} \rangle \setminus \{0\}$, then $w(x)=p^{r(e-1)}$;
\item[(iii)] $0$ otherwise.
\end{itemize}

Throughout this paper, the notation $a \equiv \Box \bmod b$ denotes that the integer $a$ is a residue quadratic modulo $b$.

\section{$\lambda$-Constacyclic Codes of Length $p^sm$ with $\lambda$ in $\F_{p^r}$ }

Let $\lambda$ be in $\F_{p^r}^*$.
In this section, we give the structure of $\lambda$-constacyclic codes of length $mp^s$ over $R$.
For this, we require the following lemma.
\begin{lem}
\label{lem:basic}
If $f(x)\in R[x]$ is a basic irreducible polynomial, then $f(x)$ is a primary polynomial.
\end{lem}
\pf Assume that $f(x)$ is basic irreducible and $g(x)h(x)\in \langle f(x) \rangle$.
%Since $f(x)$ is basic irreducible
Then $\overline{f}(x)$ is irreducible in $K[x]$, so that
$(\overline{f}(x),\overline{g}(x))=1$ or $ \overline{f}(x)$.
If $(\overline{f}(x),\overline{g}(x))=1$, then $f$ and $g$ are
coprime, and there exist $f_{1}$ and $g_{1}$ in $R[x]$ such that $1=f(x)f_{1}(x)+g(x)g_{1}(x)$.
Hence $h(x)=f(x)h(x)f_{1}(x)+g(x)h(x)g_{1}(x)$.
Since $g(x)h(x)\in \langle f(x) \rangle$, it follows that $h(x)\in \langle f(x)  \rangle$.
If $(\overline{f}(x),\overline{g}(x))=\overline{f}(x)$, then there
exist $f_1(x), g_1(x)\in R[x]$ such that $g(x)=f(x)f_1+u^{i}g_1(x)$
for some positive integer $i<e$. Then for $k>i$, we have $g^{k}\in
\langle f(x) \rangle$, and thus $f(x)$ is a primary polynomial. \qed

\begin{rem}
\label{rem:fact} Let $m$ be an integer such that $\gcd(p,m)=1$, and
$\lambda_0$ in $\F_{p^r}^*$. Then from~\cite{G-G}, the polynomial
$x^m-\lambda_0$ factors uniquely as a product of monic basic
irreducible pairwise coprime polynomials over $R$, and there is a
one-to-one correspondence between the set of monic irreducible
divisors in $\F_{p^r}$ and the basic irreducible polynomials. Since
$\F_{p^r}$ is a subring of $R$ and the decomposition of
$x^m-\lambda_0$ is unique in $R$, the polynomials $f_i$ are in
$\F_{p^r}$.
\end{rem}
\begin{lem}
\label{prop:number} Let $\alpha$ be a primitive element of
$\F_{p^r}$, and $\lambda =\alpha^{i}$ for $i \le p^r-1$. Then the
following holds:
\begin{itemize}
\item[(i)] $x^{n}= \lambda$ has a solution in $\F_{p^r}$ if and only if $(n,p^r-1)|i$;
  %where $(n,q-1)$ denotes the greatest common divisor of the integers $n$ and $q-1$.
\item[(ii)] if $n=2m$ with $m$ an odd integer and $p $ an odd prime power, then
$x^n =-1$ has a solution in $\F_{p^r}^{*}$ if and only if $-1 \equiv
\Box \bmod p^r$.
\end{itemize}
\end{lem}
\pf For Part (i), assume that $x^{n}=\lambda$ has a solution in
$\F_{p^r}$. Then this solution is equal to $\gamma=\alpha^j$ for
some $j$ and satisfies $(\alpha^{j})^n=\alpha^{i}$. This is
equivalent to $\alpha^{nj-i}=1$. Since the order of $\alpha$ is
$p^r-1$, then $(p^r-1)| nj-i \Leftrightarrow nj \,-\, r(p^r-1) = i$
for some integer $r$. This gives that $(n,p^r-1)|i$.

Assuming the existence of a solution $\alpha^i$ of $x^n=-1$, then
from Part (i) we have that $(n,p^r-1) | i$. If $n$ is even and $p$
is odd then $(n,p^r-1)$ is even, hence $i$ is even. This gives that
$-1 \equiv \Box \bmod p^r$. Conversely, assume that $-1 \equiv \Box
\bmod p^r$. Then there exists an even $i=2i'$ such that
$-1=\alpha^i$. Since $n=2m$ is oddly even,
$(-1)^m=-1=\alpha^{2mi'}=(\alpha^{i'})^n$, and hence there exists a
solution of $x^n+1=0$ in $\F_{q^{r}}$. \qed
\begin{rem}
Lemma~\ref{prop:number} gives that every $\alpha \in \F_{p^{r}}$ is
a $p^s$ power of an element in $\F_{p^{r}}$.
\end{rem}
\begin{lem}
\label{pro:deco}
Let $\lambda$ be a non-zero element of $\F_{p^r}$, and $n=mp^s$ be an integer such that $\gcd(m,p)=1$.
Then $x^n-\lambda$ has a unique decomposition over $R$ given by
\begin{equation}
\label{eq:fact} x^n-\lambda={f_1}^{p^s} \ldots {f_l}^{p^s},
\end{equation}
where the $f_i$ are irreducible polynomials coprime in $\F_{p^r}[x]$
which are divisors of $x^m-\lambda_0$, where $\lambda=\lambda_0^{p^s}$.
%Furthermore, in $\mathcal{R}[x]$
%$\langle f_i^{2^{s}} \rangle=\langle f_i^{2^{s}+l} \rangle$ for any
%$i$ and $l$, where $l$ is a positive integer.
\end{lem}
\pf From Lemma~\ref{prop:number} Part (i), we have that for any
$\lambda \in \F_{p^r}^*$ there exists $\lambda_0$ such that
$\lambda=\lambda_0^{p^s}$. That is because $\gcd(p^s,p^s-1)=1$. Hence
we have that $(x^m-\lambda_0)^{p^s}=x^{mp^s}-\lambda$ because $p|
\binom{p^s}{i}$ for $1\le i \le p^s$, and so
$x^{mp^s}-\lambda=(x^m-\lambda_0)^{p^s}$. From
Remark~\ref{rem:fact}, $x^m-\lambda_0$ has a unique decomposition
into irreducible polynomials over $\F_{p^r}$ given by
\begin{equation}
\label{eq:decom1} x^m-\lambda_0=f_1 \ldots f_l.
\end{equation}
 We need to prove that
$x^n-\lambda=f_1^{p^s} \ldots f_l^{p^s}$ is unique.
Assume that $x^n-\lambda=g_1^{\alpha_1}\ldots g_l^{\alpha_l}$ is a decomposition
into powers of basic irreducible polynomials.
From Lemma~\ref{lem:basic}, we have that the basic irreducible polynomials
are primary, hence the power of a basic irreducible polynomial is
also a primary polynomial. Then from Lemma~\ref{XIIIMac}, the
decomposition (\ref{eq:fact}) is unique.
%For the second part of the
%proposition, we have that $f_i$ and
%$\widetilde{f_i}=\frac{x^n-1}{f_i}$ are coprime, so that $f_i^l$ and
%$(\widetilde{f_i})^{2^{s}}$ are also coprime. Then there exist $g,h
%\in R[x]$ such that ${f_i}^lg+ \widetilde{f_i}^{2^{s}}h=1$. Hence
%$f_i^{l+2^{s}}g=(1-
%{\widetilde{f_i}}^{2^{s}}h){f_i}^{2^{s}}={f_i}^{2^{s}}-(x^n+1
%)^{2^{s}}h={f_i}^{2^{s}}h$. Thus in $\mathcal{R}$ we have
%that$\langle f_i^{2^{s} }\rangle=\langle f_i^{2^{s}+l} \rangle$.
\qed

\begin{prop}
\label{prop:main} With the previous notation, the primary ideals of
$\mathcal{R}$ are $\langle 0 \rangle$, $\langle 1 \rangle$, $\langle
f_i^j \rangle$, $\langle f_i^j,u^t \rangle$, with $1 \le j \le p^s$,
$1\le t < e$ and $1\le i \le l$.
\end{prop}
\pf
From~Lemma~\ref{pro:deco} we have $x^n-\lambda=f_1^{p^s} \ldots
f_l^{p^s}$. Let $\mu : R[x]\longmapsto \frac{\F_{p^r}[x]}{\langle
x^n-\lambda \rangle }$ be the canonical homomorphism.
By~Lemma~\ref{pro:deco}, we have that the factorization of
$x^n-\lambda$ over $R[x]$ is the same as that over $\F_{p^r}[x]$ and is unique.
This gives that the kernel of $\mu$ is the ideal $\langle x^n-\lambda,u\rangle$.
Hence from~\cite[Theorem 3.9.14]{Z-S}, the primary ideals of $\mathcal{R}$ are the preimages of the
primary ideals of $\F_{p^r}[x]/x^n-\lambda$.
It is well known~\cite[Theorem 3.10]{Mac2} that the primary ideals of this last
ring are the ideals $\langle f_i^j\rangle$, $1\le j\le p^s$ and $1\le i \le l$.
Hence the primary ideals of $\mathcal{R}$ are $\langle
f_i^j, u^t \rangle$.
\qed

\begin{thm}
\label{th:gen} Let $n=mp^s$ such that $(m,p)=1$, $\lambda \in
\F_{p^r}$, and $x^n-\lambda=\prod {f_i}^{p^s}$ be the unique
factorization into a product of irreducible polynomials in
$\F_{p^r}$. Then the $\lambda$-constacyclic codes of length $p^sm$
over $R$ are the ideals generated by $\langle
F_0|uF_1|\ldots|u^{e-1}F_{e-1} \rangle$, where $F_i$ for $0 \le s_i
\le F_i$ are the monic polynomial divisors of $x^n-\lambda$, and
such that $F_i |F_0$ for all $i\le e-1$.
\end{thm}
\pf
Let $\mathcal{C}$ be a constacyclic code in $R[x]$ so that $\mathcal{C}$ is an ideal of $\mathcal{R}$.
Since $\mathcal{R}$ is Noetherian, from the Lasker-Noether decomposition Theorem \cite[p. 209]{Z-S},
any ideal in $\mathcal{R}$ has a representation as a product of primary ideals.
From Proposition~\ref{prop:main}, we have that the primary ideals of $\mathcal{R}$ are $\langle f_i^j, u^t \rangle$.
Hence an ideal $I$ of $\mathcal{R}$ is of the form
\begin{equation}
\label{eq:form}  I=\prod_{l=1}^r \langle  f_i^j, u^t \rangle.
\end{equation}
Expanding the product in~(\ref{eq:form}), each
ideal in $ \mathcal{R}$ is generated by \[\langle
F_0|uF_1|u^2F_2|\ldots |u^{e-1}F_{e-1} \rangle,\] where the $F_i$ for $0
\le s_i \le e-1$ are the monic polynomial divisors of $x^n-\lambda$.
\qed
\begin{cor}
\label{cor:jit}
The cyclic codes of length $p^s$ over $R$ are of the from
\[
\langle (x-1)^{j_0}|u(x-1)^{j_1}|\ldots |u^{e-1}(x-1)^{j_{e-1}}\rangle
\]
\end{cor}
\pf Follows from Theorem~\ref{th:gen} and the decomposition $x^{p^s}-1=(x-1)^{p^s}$.
\qed
\begin{rem}
For $e=2$, the results of~Corollary~\ref{cor:jit} were also proven
in~\cite[Theorem~5.4]{dinh10} and~\cite[Theorem 3.4]{jitman}.
\end{rem}
In the following, we prove that the constacyclic codes considered in
this section are equivalent to cyclic codes.

\begin{prop}
\label{lem :2.5}
Let $n=mp^s$ such that $(m,p)=1$, and $\lambda \in \F_{p^r}$ be a $p^sm$ power in $\F_{p^r}$.
Then a $\lambda$-constacyclic code over $R$ of length $n$ is equivalent to a cyclic code of length $n$.
\end{prop}
\pf
Let $\lambda \in \F_r$ be a $p^sm$ power of $\lambda_0 \in \F_{p^r}^*$, and define
\[
\begin{tabular}{cccc}
                  % after \\: \hline or \cline{col1-col2} \cline{col3-col4} ...
               $ \phi $: $ R[x]/(x^{n}-1)$ &$ \longrightarrow$ & $R[x]/(x^{n}-\lambda)$ \\
                     $f(x) $& $\longmapsto $& $\phi(f(x))=f(\delta_0^{-1}x)$ \\
                \end{tabular}
\]
It is obvious that $\phi$ is a ring homomorphism. Hence we only need
prove that $\phi$ is a one-to-one map and it is an isometry for the
homogenous weight. For this, let $f(x)$ and $g(x)$ be polynomials in
$\F_{q}[x]$ such that $f(x)\equiv g(x) \bmod {x^n-1}.$ This is
equivalent to the existence of $h(x)\in R[x]$ such that
$f(x)-g(x)=h(x)(x^n-1)$. This equality is true if and only if
$f(\lambda_0^{-1}x)-g(\lambda_0^{-1}x)=
h(\lambda^{-1}x)[(\lambda^{-1}x)^{n}-1]$ is true. We have that
$h(\lambda^{-1}x)[\lambda^{-n}x^{n}-1]=\lambda^{-n}h(\lambda^{-1}x)[x^{n}-
\delta^{n}]=\delta^{-n}h(\delta^{-1}x)[x^{n}-\lambda]$, so for $f,g
\in \F_{q}[x]/(x^{n}-1)$
\[
\phi(f(x))\, =\, \phi (g(x)) \iff g(x)= f(x).
\]
Then $\phi$ is well defined and one-to-one, and hence is a ring
isomorphism. Now we need to prove that $\psi$ is an isometry
according to a homogeneous weight over $R$.
Let $w(.)$ be a homogeneous weight over $R$ and let $f(x)=a_0+a_1x+\ldots + a_nx^n$ be
a codeword in $R[x]/(x^{n}-1)$.
Then $\psi(f(x))=a_0+a_1\delta^{-1}x+a_2 {\delta^{-1}x}^2 +\ldots+{\delta^{-1}x}^n$.
Since $\delta$ is a unit, it must be that $w(\delta^{-i}a_i)=w(a_i)$, and hence $w(\psi(f(x))=w(f(x))$.
%As $(n,q)=1$, we know that $ R[x]/(x^{n}-1)$ and $R[x]/(x^{n}-\lambda_0)$ are principals ideal rings, Since the $\lambda$-constacyclic and
Then $\psi(A)$ is an ideal of $R[x]/(x^{n}-\lambda_0)$ and if $B$ is
an ideal of $R[x]/(x^{n}-\lambda_0)$, $\psi^{^-1}(B)$ is an ideal of
$ R[x]/(x^{n}-1)$. Since the map $\psi$ is a ring isomorphism which
is a homogeneous isometry, the codes $A$ and $\psi(A)$ are
equivalent by the result in~\cite{greferath}. \qed

\begin{thm}
\label{co:equiv4} Let $n=mp^s$ be an odd integer such that
$(m,p)=1$, and $\lambda \in \F_{p^r}^*$  such that there exists
$\delta \in \F_{q}^*$ and $\delta^{m}=\lambda$. Then the following
hold:
\begin{enumerate}
 \item [(i)]
  $\pm \lambda$-constacyclic codes of length  $m p^{s}$ over $R$ are  equivalent to cyclic codes over $R$;
  \item [(ii)]
  if $p^r \equiv 1 \bmod 4$ and $\delta =\beta^{2}$ in $\F_{q}$, then
$\pm \lambda$-constacyclic codes of length $2m p^{s}$ over $R$ are
equivalent to cyclic codes over $R$.
\end{enumerate}
\end{thm}
\pf
\begin{enumerate}
  \item [i)]
  Let $\lambda \in \F_{q}^*$ such that there exists $\delta \in \F_{q}^*$ and $\delta^{m}=\lambda$.
  %By Corollary~\ref{cor:ozbudak}
  Then there exists $\alpha \in \F_{q}^*$ such that $\alpha ^{p^{s}}$ = $\delta$,
  so that $\lambda $ = $ \delta^{m}$ = $\alpha ^{mp^{s}}$.
  Since $mp^{s}$ is odd, we obtain that $-\lambda $ = $ (-\delta)^{m}$ = $(-\alpha)^{mp^{s}}$, and the result follows
by Proposition~\ref{lem :2.5}.
  \item [ii)]
  Let $\lambda \in \F_{q}^*$ such that there exists $\beta \in \F_{q}^*$ and $\beta^{2m}=\lambda$.
  %By Corollary~\ref{cor:ozbudak}
  Then there exists $\rho \in \F_{q}^*$ such that $\rho ^{p^{s}}$ = $\beta$
  and $\lambda = \beta^{2m}= \rho ^{2mp^{s}}$. Thus $\rho$ is a $2mp^{s}$-th root of $\lambda$ in
  $\F_{q}$.
   If $q \equiv 1 \bmod 4$,
   by Lemma~\ref{prop:number} there exists $\xi \in \F_{q}$ such that $\xi^{2}=-1$.
   Then $-\lambda $ = $ (-1)^{mp^{s}} \beta^{2m}$ = $\xi^{2mp^{s}}\rho^{2mp^{s}}$ = $(\xi \rho) ^{2mp^{s}}$,
   and $\xi\rho$ is a $2mp^{s}$-th root of $-\lambda$ in $\F_{q}$.
   The result then follows from Proposition~\ref{lem :2.5}.
\end{enumerate}
\qed
\begin{ex}
For $\lambda \in \F_{p^r}^*$, the $\lambda$-constacyclic codes over
$R$ of length $p^s$ are equivalent to cyclic codes.

Let $m$ be an odd integer and $p^r \equiv 1 \bmod 4$ an odd prime power.
From Part (ii) of Theorem~\ref{co:equiv4}, the negacyclic codes of length
$2mp^{s}$ are equivalent to cyclic codes of length $2mp^{s}$ over $R$.
\end{ex}

\section{Constacyclic Codes of Length $p^sm$}

Let $\lambda=\alpha_1+u\alpha_2+\ldots + u^{e-1}\alpha_{e-1}$ be a
unit of $R$ such that $\alpha_1 \neq 0$ and $\alpha_2 \neq 0$,
and there exists $\alpha \in \F_{p^r}$ which satisfies $\alpha_1=\alpha_0^{p^s}$.
This is a generalization of the $\lambda$-constacyclic codes of Type $1^*$ given by Dinh~\cite{dinh},
thus the approach follows that in~\cite{dinh}.
Conditions on self-duality are also given.
We first prove the following lemma.
%
%$\mathcal{R}(\lambda)=\frac{R[x]}{x^{p^sm}- \lambda}$ with $\lambda$
%be a unit such that
\begin{lem}
\label{lem:p unit}
Let $p$ be an odd prime, and $\lambda=\alpha_1+u
\alpha_2+\ldots+ u^{e-1}\alpha_{e-1}$ such that $\alpha_1=\alpha_0^{p^s}\in \F_{p^r}^* $.
Then in $\mathcal{R}$ we have $(\alpha_0^{-1} x^m -1)^{p^{s}}= u\rho$ where $\rho$ is a unit
in $R$ and $\alpha_0^{-1} x^m -1$ is nilpotent in $\mathcal{R}$ with nilpotency index $ep^s$.
Furthermore, any $f\in R[x]$ which is coprime to $\alpha_0^{-1}x^m-1$ is invertible in $\mathcal{R}[x]$.
\end{lem}
\pf
In $\mathcal{R}(\alpha)$ we have
\[
\begin{array}{ccl}
(\alpha_{0}^{-1} x^m -1)^{p^s}&=&(\alpha_0^{-1} x)^{mp^{s}}+(-1)^{p^s} + \sum_{i=1}^{p^s-1}(^{p^s}_{i}) (\alpha_0^{-1} x)^{p^{s}-i}\\
&=&\alpha_0^{-p^s} x^{mp^{s}}-1 \\
&=&\alpha_1 ^{-1} ( \alpha_1 +u\alpha_2 + \ldots+
u^{e-1}\alpha_{e-1})-1 \\
&=&\alpha_1^{-1}(u\alpha_2+\ldots+ u^{e-1}\alpha_{e-1})\\
&=&u( \alpha_1^{-1}\alpha_2+\ldots+
u^{e-2}\alpha_1^{-1}\alpha_{e-1}).
\end{array}
\]
Hence $\rho=\alpha_1^{-1}(\alpha_2+\ldots+ u^{e-2}\alpha_{e-1})$ is
a unit, since $\alpha_2 \neq 0$. Thus in $\mathcal{R}$ we have
$\langle \alpha_{0}^{-1} x^m -1  \rangle= \langle u \rangle$, so the
nilpotency index of $ \alpha_{0}^{-1} x^m -1 $ is equal to $ep^s$.
For the second part of the lemma, since $f$ and $\alpha_0^{-1}x^m-1$
are coprime, there exists $g, h \in R[x]$ such that
$(\alpha_0^{-1}x^m-1)g(x)+h(x)f(x)=1$, or equivalently
$h(x)f(x)=1-Y$, where $Y=(\alpha_0^{-1}x^m-1)g(x)$. As we have
already proven that $(\alpha_0^{-1}x^m-1)^{p^{s}}=0$ in
$\mathcal{R}[x]$, we obtain that $Y^{p^{s}}=0$. Hence
$1=1-Y^{p^{s}}=(1-Y)(1+Y+\ldots+Y^{p^{s}-1})$, which means that
$h(x)f(x)$ is invertible and so $f$ is invertible in $\mathcal{R}$.
\qed

\begin{lem}
With the previous notation, in $\mathcal{R}[x]$ we have $\langle
f_i^{p^{s}} \rangle=\langle f_i^{p^{s}+k} \rangle$ for any $i$ and
$k$, where $k$ is a positive integer.
\end{lem}
\pf
We have that $f_i$ and
$\widetilde{f_i}=\frac{\alpha^{-1}x^m-1}{f_i}$ are coprime, so that
$f_i^k$ and $(\widetilde{f_i})^{p^{s}}$ are also coprime. Then there
exist $g,h \in R[x]$ such that ${f_i}^kg+
\widetilde{f_i}^{p^{s}}h=1$. Hence $f_i^{k+p^{s}}g=(1-
{\widetilde{f_i}}^{^{s}}h){f_i}^{p^{s}}={f_i}^{p^{s}}-(x^n+1)^{p^{s}}h={f_i}^{p^{s}}h$.
Thus in $\mathcal{R}$ we have that
$\langle f_i^{p^{s} }\rangle=\langle f_i^{p^{s}+k} \rangle$.
\qed
\begin{thm}
\label{th:gen}
Let $n=mp^s$ such that $\gcd(m,p)=1$, and let $\lambda=\alpha_1+u \alpha_2+\ldots+ u^{e-1}\alpha_{e-1}$
such that $\alpha_1=\alpha_0^{p^s}\in \F_{p^r}^* $ and $\alpha_2 \neq 0$.
Then $x^n-\alpha_1=\prod {f_i}^{p^s}$ factors uniquely as the product of
irreducible polynomials in $\F_{p^r}$ and $ \mathcal{R}$ is a
principal ideal ring whose ideals are generated by
\begin{equation}
\label{eq:gen}
 \langle \prod f_i ^{s_i}\rangle, \text{ where }0 \le
s_i \le p^{s}e.
\end{equation}
Moreover we have that
\[
|\mathcal{C}|= p^{re(n- \sum _{i \in I} s_i \deg f_i )}.
\]
\end{thm}
\pf  Since $\alpha_0^{p^s}=\alpha_1$, from~Lemma~\ref{pro:deco} we have that $x^n-\alpha_1=f_1^{p^s} \ldots
f_l^{p^s}$ factors uniquely as the product of irreducible polynomials in $\F_{p^r}$.
Let $\mathcal{C}$ be a $\lambda-$constacyclic code in $R$ so that $\mathcal{C}$ is an ideal of $\mathcal{R}$.
Further, let $\mathcal{C}_{u}$ be the set of codewords of $\mathcal{C}$ reduced modulo $u$.
Then $\mathcal{C}_{u}$ is an ideal of $\frac{\F_{p^r}[x]}{x^n-\alpha_1}$, and hence
$\mathcal{C}_{u}=\langle \prod {f_i}^{l_i} \rangle$, where $l_i \le p^s$.
Then for $c(x)\in \mathcal{C}$, there exists $g$ and $h$ in
$\mathcal{R}$ such that $c(x)=h(x)\prod {f_i}^{l_i}(x)+ug(x)$.
However, since in $\mathcal{R}_{\lambda}$ we have from
Lemma~\ref{lem:p unit} that $ u\in \langle (\alpha_0^{-1}x^m-\alpha_1)^{p^s} \rangle= \prod {f_i}^{p^{s}}(x)$,
then $\mathcal{C} \subset \langle \prod {f_i}^{l_i}(x) \rangle$.
Let $s_i$ be the largest power of $f_i$ such that $\mathcal{C} \subset \langle \prod {f_i}^{s_i}(x) \rangle$.
Then from Proposition~\ref{pro:deco}, we have $0 \le s_i \le p^{s}$.
If $c(x)\in \mathcal{C}$, then $c(x)=c'(x)\prod {f_i}^{s_i}(x)$ with $c'(x) \in \mathcal{R}$.
Since the $s_i$ are maximal, we have that $\gcd(c'(x),\prod {f_i}^{s_i}(x))=1$, and hence $\gcd(c'(x,
\alpha_0^{-1}x^m-1)=1$.
By Lemma~\ref{lem:p unit}, $c'(x)$ is invertible, so that $\prod {f_i}^{s_i}(x)=c'(x)c^{-1}(x) \in \mathcal{C}$.
Thus $\mathcal{C}=\prod {f_i}^{s_i}(x)$ with $s_i \le p^{s}e$,
and hence $|\mathcal{C}|=|R|^{n-\sum s_i \deg (f_i)} = p^{re(n-\sum s_i \deg (f_i))}$.
\qed

Since the $f_i$ are pairwise coprime, the ideals $\langle \prod_{i \in I} f_i^{s_i}\rangle$, $0 \le s_i \le p^{s}e$, are distinct.
Hence the number of $\lambda$-constacyclic codes is given in the following lemma.
\begin{cor}
With the previous notation, the number of $\lambda$-constacyclic codes of length $mp^s$ is equal to
\[
(p^{s}e+1)^{l},
\]
where $l$ is the number of the $p^r$-cyclotomic classes modulo $m$.
\end{cor}
\begin{ex}
If $m=1$, then $x^{p^s}-\alpha_1=(x-\alpha_0)^{p^s}$. Hence in this
case $\mathcal{R}$ is a chain ring with the following ideals
\[
\langle 0 \rangle= \langle (x-\alpha_0)^{ep^s}) \rangle \subsetneq
\langle (x-\alpha_0)^{ep^s-1}) \rangle \ldots  \subsetneq \langle
(x-\alpha_0) \rangle \subsetneq \langle 1 \rangle =\langle
\mathcal{R}\rangle .
\]
\end{ex}
\subsection{Self-Dual $\lambda$-Constacyclic Codes}
In this section, we study the self-duality of $\lambda$-constacyclic codes.
\begin{thm}
\label{thm:self}
Let $n=mp^s$ such that $\gcd(m,p)=1$, and $\lambda=\alpha_1+u \alpha_2+\ldots+ u^{e-1}\alpha_{e-1}$
such that $\alpha_1=\alpha_0^{p^s}\in \F_{p^r}^* $ and $\alpha_2 \neq 0$.
Let $\mathcal{C}$ be a $\lambda$-constacyclic code of length $p^sm$ generated by the polynomial
$\prod_{i} f_i^{s_i}$ with $s_i \le p^se$.
Then the dual $\mathcal{C}^\bot$ of $\mathcal{C}$ is a $\lambda^{-1}$-constacyclic code of length $mp^s$.
If $\lambda^2=1$, then $\mathcal{C}^{\bot}$ is a $\lambda$-constacyclic code generated by
\[
\prod_{i}(f_i^*)^{p^{s}e-s_i} .
\]
\end{thm}
\pf We know that the dual of a $\lambda$-constacyclic code
$\mathcal{C}$ is a $\lambda^{-1}$-constacyclic code. Hence
$\mathcal{C}$ can be self-dual if and only if $\lambda^2=1$.
From~(\ref{wood}), the cardinality is
$|\mathcal{C}^{\bot}|=\frac{|R|^n}{p^{re(n-\sum s_i \deg
(f_i))}}=p^{re\sum s_i \deg (f_i))}$. Now denote
$\Hat{\mathcal{C}}=\langle \prod_{i}(f_i)^{p^se-s_i} \rangle$. Then
we need only prove that $\Hat{\mathcal{C}}^* \subset
\mathcal{C}^{\bot}$ have the same cardinality. We have the following
equality in $\mathcal{R}$
\[
\prod_{i}(f_i)^{p^{s}e-s_i}\prod_{i}(f_i)^{s_i}=\prod_{i}(f_i)^{p^se}=(x^m-\alpha_1)^{p^se}=0.
\]
Hence it follows from Lemma~\ref{sel:dinh} that
$\Hat{\mathcal{C}}^*\subset ann^*(\mathcal{C})=\mathcal{C}^{\bot}$.
Since $\Hat{\mathcal{C}}$ is an ideal of $\mathcal{R}$, from
Theorem~\ref{th:gen} we have that
$|(\Hat{\mathcal{C}})^*|=|\Hat{\mathcal{C}}|= p^{re(\sum _{i \in I}
s_i \deg f_i )}$. This has the same cardinality as
$\mathcal{C}^{\bot}$, since
$\mathcal{C}^{\bot}=\frac{|R^n|}{p^{re-(n\sum _{i \in I} s_i \deg
f_i )}}=p^{re(\sum _{i \in I} s_i \deg f_i )}$ from (\ref{wood}).
\qed

Now we give condition on the existence of $\lambda$-constacyclic self-dual codes over $R$.
For this we need the following decomposition.

Denote the self-reciprocal factors in the factorization of
$x^m-\alpha_0$ by $g_1, \ldots, g_k$, and the remaining factors
grouped in pairs by $h_1,h_1^\star,\ldots, h_t,h_t^\star$. Hence
$l=k+2t$, and we have the following factorization:
\begin{equation}
\begin{array}{ccl}
\label{eq:ling}
x^n-\alpha_1&=&(x^m-\alpha_0)^{p^s}=g^{p^s}_1(x) \ldots g^{p^s}_k(x)\\
&&\times h^{p^s}_1(x)h^{\star p^s}_1(x) \ldots h^{p^s}_t(x)h^{\star
p^s}_t(x).
\end{array}
\end{equation}
\begin{thm}
\label{th:exist} Let $p$ be an odd prime with the notation above,
$\lambda$ such that $\lambda^2=1$, and $e$ odd. Then there exists a
self-dual $\lambda$-constacyclic code of length $mp^s$ over $R$ if
and only if $m$ is even and there are no $g_i$ (self-reciprocal
polynomials) in the factorization of $x^{mp^s}-\alpha_1$ given
in~(\ref{eq:ling}). Further, a self-dual $\lambda$-constacyclic code
$C$ is generated by a polynomial of the form
\begin{equation}
h^{b_1}_1(x)h^{\star p^se- b_1}_1(x) \ldots h^{b_t}_t(x)h^{\star p^se-b_t}_t(x).
\end{equation}
\end{thm}
\pf From Remark~\ref{rem:wood}, we must have that $m$ is even since
it was assumed that $e$ and $p$ are odd. If there exists a
$\lambda$-constacyclic self-dual code $\mathcal{C}$ of length
$n=mp^s$ over $R$, then from (\ref{eq:gen}) it is generated by
$A(x)= \prod {f_i}^{k_i},$ where the $f_i$ are factors of
$x^m-\alpha_0$. From~(\ref{eq:ling}), we can write
\[
A(x)=g^{a_1}_1(x)\ldots g^{a_k}_k(x)h^{b_1}_1(x)h^{\star
c_1}_1(x)\ldots h^{b_t}_t(x)h^{\star c_t}_t(x),
\]
where $0 \le a_i \le p^se$ for $ 1 \le i \le k$, and $0 \le b_j \le
p^se$ and $0 \le c_j \le p^se$ for  $ 1 \le j \le t$. Let $\langle
B(x) \rangle=\mathcal{C}^{\bot}$.
Then from~Theorem~\ref{thm:self} we obtain
\[
\begin{array}{ccl}
B(x)&=&g^{p^se-a_1}_1(x)\ldots g^{p^se-a_k}_s(x)\\
&&\times h^{\star p^se-b_1}_1(x)h^{ p^se-c_1}_1(x)\ldots
h^{\star p^se-b_t}_t(x)h^{ p^se- c_t}_t(x).
\end{array}
\]
Since $\mathcal{C}$ is self-dual, by equating factors of $A(x)$
and $B(x)$, the powers of the factors of $A(x)$ and $B(x)$ must
satisfy $a_i = p^se-a_i$ for $1\le i \le k$, and $c_j= p^se- b_j$ for $1 \le j \le t$.
Equivalently, $p^se =2a_i$ for $1\le i \le k$, and $c_j=p^se- b_j$ for $1 \le j \le t$.
Since $pe$ is odd, the last equalities are possible if and only if there are no $g_i$
in the factorization of $x^{mp^s}-\lambda$ and $c_j= p^se- b_j$ for $1 \le j \le t$,
i.e, $k=0$ in (\ref{eq:ling}) and $c_j= p^se- b_j$ for $1 \le j \le t$.
Hence a $\lambda$-constacyclic self-dual code is generated by
\[
h^{b_1}_1(x)h^{\star p^se-b_1}_1(x)\ldots h^{b_t}_t(x)h^{\star p^se-b_t}_t(x).
\]
\qed

The condition given in Theorem~\ref{th:exist} that there are no
irreducible factors of $x^m-\alpha_0$ which are self-reciprocal is
equivalent to $p^{ri} \neq -1 \mod m$ for all positive integers
$i$~\cite{G-G122}. Hence we obtain the following corollary.
\begin{cor}
\label{cor:exists}
With the previous notation, assume that $\lambda$ is such that
$\lambda^2=1$, $p$ is an odd prime, $e$ is odd and $m$ is an even integer.
Then non-trivial $\lambda$-constacyclic self-dual codes of
length $n$ over $R$ exist if and only if $(p^r)^i\neq -1 \pmod m$
for all positive integers $i$.
\end{cor}

\begin{cor}
With the previous notation, assume that $\lambda$ is such that
$\lambda^2=1$, $e$ is odd, $m$ an even integer, and $p$ an odd
prime such that $p\equiv 5 \mod 8$ or $p \equiv 3 \mod 8$.
Then no $\lambda$-constacyclic code exists over $R$.
\end{cor}
\pf From~\cite[Theorem 6]{moree}, if $p\equiv 5 \mod 8$ or $p \equiv
3 \mod 8$, then there exists $i >0$ such that $(p^r)^i\neq -1 \pmod
m$. The result then follows from Corollary~\ref{cor:exists}. \qed
\begin{ex}
If $e$ is odd. Then there is no $\lambda$-constacyclic self-dual
code of length $5^s\cdot6,$ $s\ge 1$ over $\F_5+u\F_5+\ldots
+u^{e-2}\F_5$.
\end{ex}

\end{document}